\documentstyle{article}
\begin{document}
\renewcommand{\baselinestretch} {1.5}
\large
\hyphenation{anti-ferro-mag-netic}

\begin{flushright}
{\large{\bf YAMANASHI-94-02\\ Octorber 1994}}
\end{flushright}
\vskip 1.0in
\begin{center}
{\large{\bf A Hybrid Method for Global Updates\\
in Monte Carlo Study\\}}
\vskip 1.0in
   Tomo Munehisa and Yasuko Munehisa\\
\vskip 0.5in
   Faculty of Engineering, Yamanashi University\\
   Kofu, Yamanashi, 400 Japan\\
\vskip 1.5in
{\bf Abstract}
\end{center}

We propose a new algorithm which works effectively in global updates
in Monte Carlo study. We apply it to the quantum spin chain with
next-nearest-neighbor interactions. We observe that Monte Carlo results
are in excellent agreement with numerically exact ones obtained by the
transfer matrix method.

\eject

The Monte Carlo method has been an indispensable tool for theoretial
study in physics. An essential point of the method is to generate
configurations according to their Boltzmann weights. For this purpose
one usually starts with an initial configuration and obtains a new
configuration sequantially by some rule that guarantees the
ergodicity and satisfies the condition of the detailed balance.
Widely used rules are the heat bath method and the Metropolis
algorithm. The heat bath mehtod is useful for systems which have
a small number of candidates for updating because one has to
calculate weights for all candidates to determine which one should
make the new configuration. The Metropolis algorithm is suitable for
complicated systems since in this algorithm it is enough to know the
weight just for one candidate to decide whether one should accept or
reject that candidate for updating. The problem in the latter is that
the efficiency of updating would be bad if the selection of the
candidate is not appropriate.

In previous papers\cite{mune,mune2} for quantum Monte Carlo study
using the Suzuki-Trotter formula we proposed re-strucuring
method to improve the negative sign problem and applied it to one
dimensional spin-1/2 system with next-nearest-neighbor
interactions. It turned out that this method is very useful for
this model. We found however that it was difficult to make
Monte Carlo simulations with large values of the Trotter number
for technical reasons on the so-called global updates, which
are necessary in the quantum Monte Carlo method in addition to
the usual local updates.
In the global updates one needs to change spin states on all sites
along some lines connecting these sites, which usually depend the
size of the system, while
in the local updates limited part of the system
irrelevant to whole size of the system should be updated.
In the conventional apporach to the model, where
the state on each site is represented spin up or spin down,
conservation property of the quantum number $J_z$ ($z$-component
of the total spin $J$) is helpful to carry out global updates
for large sizes of the system. In the re-structuring approach,
on the contrary, we employ not eigenstates of $\sigma_i^z$, where
$\sigma_i^z$ denotes $z$-component of the Pauli matrix on site
$i$, but eigenstates of ${\vec \sigma}_{2i-1} {\vec\sigma}_{2i}$,
that is $1_i, \oplus_i, \ominus_i or -1_i$.
They are eigenstates of $\sigma_{2i-1}^z+\sigma_{2i}^z$ with
eigenvalues $2,0,0,-2$. Because of two $J_z=0$ states
we need to consider $\sim 2^{2n}$
candidates for a global update in the Trotter direction with
the Trotter number $n$, which causes shortage of computer
memory when $n$ becomes large. Even for small values of $n$ we found
the Metropolis algorithm gives too poor efficiency to obtain results.
In previous Monte Carlo work we therefore simulated systems with
the Trotter number $n \leq 4$ only by employing the heat bath
method with an approximation to discard $\Delta J_z = 2$
candidates. The agreement between Monte Carlo data and
numerically exact calculations are satisfactory, but small
discrepancy remains.
In this paper we propose hybrid use of the heat bath method and the
Metropolis algorithm which is applicable for large values of the
Trotter number. We will see the discrepancy mentioned above
is completely resolved.

Now let us discribe the hybrid method. In this method we choose
a candidate state applying the heat bath method repeatedly and use
the Metropolis algorithm to determine whether we accept it or not.
For concrete description we consider the quantum spin-$1/2$ system
on an $N$-sited chain with next-nearest-neigbor interactions.
After the re-structuring the system reduces to an effective
quantum spin system sitting on an $N/2$-sited chain
which has nearest-neigbor interactions only.
Along the Trotter direction
we have $2n$ sites with the Trotter number $n$. Each site $(i,j)$
on such two-dimensional lattice, where $i(j)$ denotes the location
in the space (Trotter) direction, is characterized by a spin
state $S_{(i,j)}$ with $S=1, \oplus, \ominus$ or $-1$.
For one $n$-direction global updates with a fixed $i$ we have to
find a set of $S_{(i,j)}$, $(j=1,2,...,2n)$, as a candidate.
Let us denote $S_{(i,j)}$ in the old configuration as $\alpha_j$
and its candidate as $\beta_j$ hereafter.
First we consider the plaquette $Q_1$ in Fig.1(a) only,
where it is easy to pick up $\beta_1$ and $\beta_2$ according to
the probability $P^{(1)}$,
$$ P^{(1)}=W(\gamma_1,\beta_1,\gamma_2,\beta_2) /
 \sum_{\beta_1} \sum_{\beta_2}
W(\gamma_1,\beta_1,\gamma_2,\beta_2),$$

\noindent
with $\gamma_j$ being the state of the nearest-neighbor site of
the site $(i,j)$ and

\noindent
$W(\zeta_a,\zeta_b,\zeta_c,\zeta_d)$ being
the Boltzmann weight for a plaquette depicted in Fig.1(b)\footnote{We
have $4^4=256$ values of W in this model. Actually,
only $62$ of them are non-zero.}.
Next we choose $\beta_3$ in the plaquette $Q_2$ according to the
probability $P^{(2)}$,
$$ P^{(2)}=W(\beta_2,\gamma_2,\beta_3,\gamma_3) /
 \sum_{\zeta_3} W(\beta_2,\gamma_2,\zeta_3,\gamma_3).$$

\noindent
Similarly we choose $\beta_{j+1}$ ($j=3,4,...,2n-1$) sequentially
according to the probability
$P^{(j)}$ which is, for odd $j$,
$$ P^{(j)}=W(\gamma_j,\beta_j,\gamma_{j+1},\beta_{j+1}) /
 \sum_{\zeta_{j+1}} W(\gamma_j,\beta_j,\gamma_{j+1},\zeta_{j+1}) $$

\noindent
or, for even $j$,
$$ P^{(j)}=W(\beta_j,\gamma_j,\beta_{j+1},\gamma_{j+1}) /
 \sum_{\zeta_{j+1}} W(\beta_j,\gamma_j,\zeta_{j+1},\gamma_{j+1}).$$

\noindent
Finally we compare the probability $P_{\rm new}$ and $P_{\rm old}$
$$ P_{\rm new}=W(\beta_{2n},\gamma_{2n},\beta_1,\gamma_1) $$
$$\times \prod_{k=1,n-1}
\sum_{\zeta_{2k+1}} W(\beta_{2k},\gamma_{2k},\zeta_{2k+1},
\gamma_{2k+1}) \sum_{\zeta_{2k+2}} W(\gamma_{2k+1},\beta_{2k+1},
\gamma_{2k+2},\zeta_{2k+2}), $$

$$ P_{\rm old}=W(\alpha_{2n},\gamma_{2n},\alpha_1,\gamma_1) $$
$$\times \prod_{k=1,n-1}
\sum_{\zeta_{2k+1}} W(\alpha_{2k},\gamma_{2k},\zeta_{2k+1},
\gamma_{2k+1}) \sum_{\zeta_{2k+2}} W(\gamma_{2k+1},\alpha_{2k+1},
\gamma_{2k+2},\zeta_{2k+2}), $$

\noindent
and decide, following the Metropolis algorithm,
whether we replace $\alpha_1,\alpha_2,...,\alpha_{2n}$
by $\beta_1,\beta_2,...,\beta_{2n}$ or not.
It is easy to see that this method satisfies the condition
of the detailed balance.
In Fig.2 we plot Monte Carlo results on the $R$ ratio obtained by
this method, together with numerical exact calculations by the
transfer matrix method. We perform simulations on an $N=8$ chain
with the Trotter number up to $16$. We observe that
the Monte Carlo data and the results from the transfer matrix
method almost coincide for every value of $n$. It should be also
noticed that the agreement is remarkable even for large values of
the inverse temperature $\beta$ because of much better efficiency
of the updating than the efficiency in the previous heat bath method.

To summarize we propose a hybrid method for global updates in the
Monte Carlo algorithm. Through numerical work for a quantum spin
system we show the method is quite powerful.
Since the procedure in this method is quite general we expect
it is applicable to many systems in the quantum Monte Carlo
studies. We also expect the method is helpful in the classical
statistical mechanics with the cluster algorithm\cite{clust},
where the global updates become important especially when the
correlation length is large.
We would like to emphasize that the hybrid method enables us
to choose appropriate candidates for the update at only a little
cost of the computer time and the memory.

\vfill
\eject

\vskip 1.0in

\noindent {\bf Figure Captions}

\noindent {\bf Figure 1}\\
(a) A schema of a part of the two-dimensional lattice with
the Trotter number $n=4$ to show how the candidate states are
chosen for the $n$-direction global update in the hybrid method.
The horizontal (vertical) arrow denotes the space (the Trotter)
direction.\\
(b) A plaquette with sites $a$, $b$, $c$ and $d$.
The arrows are the same as in Fig.(a).

\noindent {\bf Figure 2}\\
The $R$ ratio, the ratio between $(Z_+ - Z_-)$ and $(Z_+ + Z_-)$
of the re-structured quantum spin system calculated by the Monte
Carlo simulations with the hybrid method (mc) and by the transfer
matrix method (tm) on an $N=8$ chain. Here $Z_+(Z_-)$ denotes number
of positively (negatively) weighted configurations generated in
the simulation. Exact values of $R$ are also
plotted for comparison. Monte Carlo results in the figure
with the Trotter number $n=2,4,8$ ($n=16$) is an average of
a dozen of data calculated from 10000 (100000) configurations after
2000 (20000) sweeps for thermalization. The errors are statistical
ones only.

\vfill

\end{document}